\documentclass[12pt]{article}
\pdfoutput=1
\usepackage{graphicx,psfrag,epsf,color}
\usepackage{amsmath,amssymb,amsfonts}
\usepackage{array}
\usepackage{bm} 
\usepackage{cite}
\usepackage{rotating}
\usepackage{slashed, cancel}
\usepackage{xcolor}
\newcounter{MBQ}

\newcounter{RSQ}

\newcommand{\be}{\begin{equation}}
\newcommand{\ee}{\end{equation}}
\newcommand{\bea}{\begin{eqnarray}}
\newcommand{\eea}{\end{eqnarray}}
\newcommand{\bi}{\begin{itemize}}
\newcommand{\ei}{\end{itemize}}
\newcommand{\ben}{\begin{enumerate}}
\newcommand{\een}{\end{enumerate}}
\newcommand{\bt}{\begin{tabular}}
\newcommand{\et}{\end{tabular}}

\newcommand{\mchi}{m_\chi}

\begin{document}
\allowdisplaybreaks

\begin{titlepage}

\begin{flushright}
{\small
TUM-HEP-1222/19\\
September 09, 2019\\
}
\end{flushright}

\vskip1cm
\begin{center}
{\Large \bf\boldmath Wino potential and Sommerfeld effect 
at NLO}
\end{center}

\vspace{0.5cm}
\begin{center}
{\sc Martin~Beneke,$^a$ Robert~Szafron,$^a$ Kai Urban$^{a}$} \\[6mm]
{\it $^a$Physik Department T31,\\
James-Franck-Stra\ss e~1, 
Technische Universit\"at M\"unchen,\\
D--85748 Garching, Germany
}
\end{center}

\vspace{0.6cm}
\begin{abstract}
\vskip0.2cm\noindent
We calculate the SU(2)$\times$U(1) electroweak static potential 
between a fermionic triplet in the broken phase of the 
Standard Model in the one-loop order (NLO). The one-loop correction 
provides the leading non-relativistic correction to the 
large Sommerfeld effect in the annihilation of wino or 
wino-like dark matter particles~$\chi^0$. We find sizeable 
modifications of the $\chi^0\chi^0$ annihilation cross section 
and determine the shifts of the resonance locations due to the 
loop correction to the wino potential.
\end{abstract}
\end{titlepage}

\section{Introduction}
\label{sec:Introduction}

It is by now well-known that the Sommerfeld effect due to the 
electroweak Yukawa force \cite{Hisano:2003ec,Hisano:2004ds,Hisano:2006nn} can lead 
to a dramatic enhancement of the annihilation cross section of 
two dark matter (DM) particles if their mass is in the TeV range. 
Contrary to the classic Sommerfeld effect for massless gauge 
boson exchange in QED and QCD, which rises as $1/v$ as the relative 
velocity of the annihilating particles decreases, the enhancement 
due to the Yukawa force saturates at small velocities, except near 
isolated resonances. These occur at dark matter mass values, when a 
zero-energy bound-state develops in the spectrum. The phenomenon is 
quite general and also appears for lighter DM, if there is a 
force carrier with even smaller mass~\cite{ArkaniHamed:2008qn}. 
Furthermore, if the DM particle is part of a multiplet with a 
small mass splitting, the effect depends sensitively on the mass 
difference \cite{Slatyer:2009vg}. 

Its main interest is nevertheless due to the fact that it is a 
generic feature of the classic WIMP DM particle, where it arises 
from the well-established Standard Model (SM) interactions. Thus, 
it appears in the so-called minimal models \cite{Cirelli:2007xd} 
and for TeV scale MSSM WIMPs (see, for example, 
\cite{Hryczuk:2010zi,Beneke:2014gja,Beneke:2014hja}). The 
Sommerfeld effect is particularly important for the annihilation 
rates and relic density of the pure wino, an electroweak triplet 
of fermions of which the electrically neutral member is the DM 
particle \cite{Hisano:2004ds,Hisano:2006nn,Cirelli:2007xd,Hryczuk:2011vi,Fan:2013faa,Cohen:2013ama}, or a mixed but dominantly wino 
state \cite{Beneke:2016ync,Beneke:2016jpw}. The pure wino 
(``wino'' in the following) model has become a test case for 
the quantitative understanding of large electroweak corrections 
in the annihilation of TeV scale DM particles.
In view of the possible detection or exclusion of the wino 
particle through measurements of high-energy cosmic rays, 
the wino annihilation rate into photons is of particular 
interest. Here electroweak perturbation theory breaks down 
due to electroweak Sudakov logarithms, which must be summed 
in addition to the Sommerfeld corrections. Recent work on 
exclusive and semi-inclusive photon yields has shown that 
Sudakov logarithms can be controlled with 1\% accuracy with 
NLL' resummation \cite{Ovanesyan:2016vkk,Beneke:2018ssm,Beneke:2019vhz}. At this level of precision, the treatment of the Sommerfeld 
effect should be revisited, since, up to the present, all 
calculations have been done with the tree-level exchange potential, 
which corresponds to the leading-order (LO) approximation 
in non-relativistic effective field theory (EFT) for the 
DM particle \cite{Beneke:2012tg,Hellmann:2013jxa,Beneke:2014gja}.

In this paper we compute the one-loop corrections to the 
wino potential and discuss its effect on the wino pair annihilation 
cross section to photons, $\chi^0\chi^0\to\gamma+X$. We 
recall  \cite{Hisano:2003ec,Hisano:2004ds,Hisano:2006nn} 
that the LO potential is given by the matrix 
\begin{equation}
V_{\rm LO}(r)   = 
  \left(\begin{array}{cc}\displaystyle 
0 &  \displaystyle 
\quad - \sqrt{2}\,\alpha_2 \,\frac{e^{-m_W r}}{r}  
\\
\displaystyle   - \sqrt{2}\,\alpha_2 \,\frac{e^{-m_Wr}}{r} & 
\displaystyle \quad  -\frac{\alpha}{r} 
- \alpha_2 \,c_W^2 \,\frac{e^{-m_Z r}}{r} 
  \end{array}\right).
 \label{eq:VLO}
\end{equation}
The $IJ$ entry refers to the non-relativistic scattering 
of wino two-particle states $I\to J$ with $I, J=1,2$ referring 
to $\chi^0\chi^0$ and $\chi^+\chi^-$, respectively. The 
above matrix describes the scattering of electrically neutral 
two-particle states in a ${}^1S_0$ spin-angular-momentum 
configuration, since the spin-1 configuration is forbidden 
due to the Majorana nature of the $\chi^0$. One might expect 
the one-loop correction to the potential to be small due to 
the smallness 
of the SU(2)$\times$U(1) couplings. However, we shall see 
that over most of the interesting wino mass range from 
1 to 10~TeV, the effect on the annihilation cross section 
is significantly larger than the typical $3\%$ of 
an electroweak quantum effect. 

In the following we give only a brief overview of technical 
details of the computation and then present results for 
the potential and the annihilation cross section 
into photons. An NLO Sommerfeld calculation of the relic 
density involves the potentials for all coannihilation 
channels. We leave this to a longer and more technical paper.

\section{Technical details}
\label{sec:technical}

The Sommerfeld effect is a low-energy phenomenon that appears for 
non-relativistic DM particles. A systematic treatment of 
non-relativistic effects can be given in non-relativistic and 
potential-non-relativistic DM 
EFT~\cite{Beneke:2012tg,Beneke:2014gja}. 
The potential appears in the effective Lagrangian 
\begin{eqnarray}
\label{eq:LPNRDM}
\mathcal{L}_{\rm PNRDM} &=& \sum_i \chi_{vi}^\dagger(x) \left(i D^0(t,\mathbf{0}) - 
\delta m_i+ \frac{\bm{\partial}^2}{2\mchi}\right)\chi_{vi}(x) 
\nonumber\\
&&-\,\sum_{\{i,j\},\{k,l\}}
\int d^3\mathbf{r}\, V_{\{ij\}\{kl\}}(r)\,
\chi_{vk}^\dagger(t,\mathbf{x})\chi_{vl}^\dagger(t,\mathbf{x}+\mathbf{r}) 
\chi_{vi}(t,\mathbf{x})\chi_{vj}(t,\mathbf{x}+\mathbf{r}) 
\quad
\end{eqnarray}
as an instantaneous but spatially non-local interaction of 
four non-relativistic wino fields $\chi_{vi}$ where 
$i=0,+,-$.\footnote{This defines the potential as a 
$3\times 3$ matrix for the two-particle states 
$ij=00,+-,-+$ in the sector with zero electric charge. This 
is the most general definition which automatically takes 
care of the (anti-)symmetrization properties. 
For practical applications it is more conventional to remove 
the redundant $-+$ state, to 
project the potential on channels with given spin and 
angular momentum, and to work with the $2\times 2$ matrix 
in the space of two-particle states, see (\ref{eq:VLO}).  
The relation between the two conventions is explained in 
Section~3 of \cite{Beneke:2014gja}. In the following 
we work with the $2\times 2$ matrix formalism (method-2 in 
 \cite{Beneke:2014gja}).} $\delta m_i$ denotes the small 
mass splitting between the $\chi^-$ and the $\chi^0$ state.

Standard non-relativistic power counting for the wino 
assumes $\alpha_2\sim v\sim m_W/m_\chi$, although it is then 
possible to consider $v\ll \alpha_2$.  The potential generated 
by tree-level gauge boson exchange is then a leading-order  
interaction -- as large as the kinetic term. Treating this 
interaction as part of the unperturbed Lagrangian and 
solving the corresponding Schr\"odinger equation gives the 
LO Sommerfeld effect. Similarly, the radiative mass splitting 
$\delta m_i \sim m_W \alpha_2$ at the one-loop order is of 
the same order as $\partial_0 
\sim E \sim m_\chi v^2$, and therefore relevant at LO. 
NLO corrections, that is, corrections suppressed by one 
power of $\alpha_2$, $v$ or $m_W/\mchi$ to the above 
Lagrangian arise from 
a) the two-loop correction to the mass splitting, which is 
known \cite{Yamada:2009ve,Ibe:2012sx}, b) the one-loop
correction to the Yukawa/Coulomb potential (\ref{eq:VLO}), 
which is the subject of this paper, and, possibly from 
c) potentials with more singular short-distance behaviour 
than $1/r$, similar to the massless gauge boson case, and 
d) ultrasoft gauge-boson radiation. However, the latter two 
effects do not appear at NLO for the same reason as in 
QCD and QED. Note that there exist of course NLO corrections 
to the annihilation process (see, for example,  
\cite{Baumgart:2017nsr,Beneke:2018ssm,Beneke:2019vhz}), 
but here we are concerned with non-relativistic effects.

The potential is technically a matching coefficient between 
non-relativistic and potential non-relativistic DM EFT. It is 
obtained from the wino-wino scattering amplitude 
$i\,T^{\chi\chi \to \chi\chi}_{ijkl}(\bm{q})$
at small momentum transfer $\bm{q}$. At the one-loop 
order, the matching coefficient is extracted from the 
soft region in the method-of-region expansion \cite{Beneke:1997zp}, 
which is automatic, if one replaces the non-relativistic 
wino propagators by static propagators $i/p^0$, and picks 
up the poles in the loop-momentum zero-component $k^0$ 
from the gauge-boson propagators. The coordinate-space potentials 
follow by taking the Fourier transform 
\begin{align}
V_{ \lbrace ij\rbrace \lbrace kl\rbrace }(r) = 
\int \frac{d^3\bm{q}}{(2\pi)^3} \, e^{i\bm{q}\cdot \bm{x}} \,
i\,T^{\chi\chi \to \chi\chi}_{ijkl}(\bm{q}^{2})
\,,
\label{eq:fourier}
\end{align}
where $r\equiv|\bm{x}\,|$. From the identity 
\begin{align}
\int \frac{d^3\bm{q}}{(2\pi)^3} \, e^{i\bm{q}\cdot \bm{x}} \,
\frac{1}{\bm{q}^{2}+m^2} = \frac{e^{-m r}}{4\pi r}
\,,
\label{eq:yukawa}
\end{align}
one recognizes the well-known Yukawa-like potential for amplitudes 
with exchange of a force carrier of mass $m$. 

\begin{figure}[t]
\centering
\includegraphics[width=0.8\textwidth]{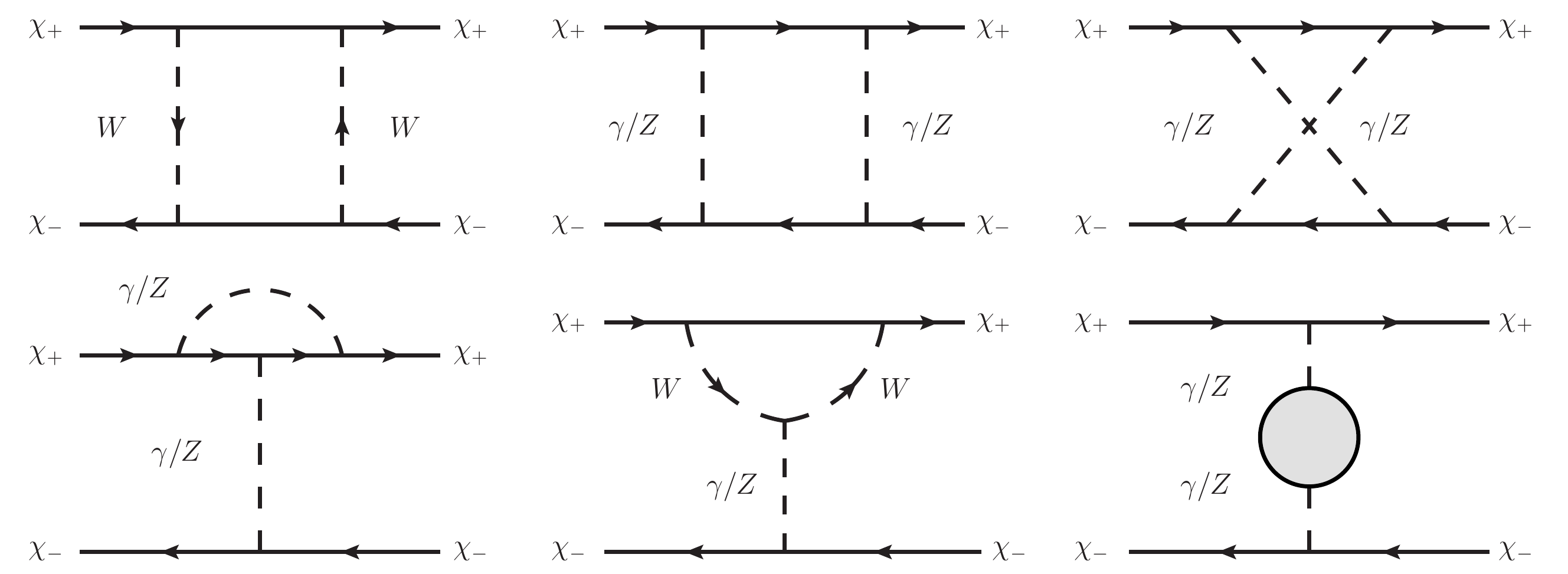}
\caption{Feynman diagrams for the $\chi^+ \chi^- 
\to \chi^+ \chi^-$ scattering channel 
(excluding field renormalization, counterterm and tadpole diagrams). 
Arrows on propagators indicate charge flow. For the 
$\chi^0 \chi^0 \to \chi^0 \chi^0$ channel, 
of the above diagrams only the box and crossed 
box with $W$-exchange exist and cancel against each other. 
In the $\chi^0 \chi^0 \to \chi^+ \chi^-$ channel, the same 
topologies as above are possible (with different bosons such that 
charge flow is respected), except for the crossed box diagram.}
\label{fig:1looppmpm}
\end{figure}

Following this procedure, the calculation of the one-loop 
correction to the wino potential is standard, and involves 
the Feynman diagrams shown in Figure~\ref{fig:1looppmpm}.  
We performed the calculation in general covariant gauge with 
different gauge parameters $\xi$ for the $W$-, $Z$-boson and 
photon, and find that the result does not depend on the gauge-fixing 
parameters, as required.\footnote{The tadpole diagrams require the 
standard electroweak treatment and as expected do not affect the 
final result \cite{Denner:1991kt}. However, it is useful to keep 
track of them, as they make the coupling and mass counterterms 
separately gauge invariant \cite{Fleischer:1980ub}. We also 
note that the diagram involving the triple gauge-boson vertex 
vanishes in Feynman gauge, but does not in other gauges.}
The diagrams are reduced to a few master integrals, 
which are then calculated analytically. For the gauge boson 
self-energy diagrams in general covariant gauge we used 
\texttt{FeynArts} \cite{Hahn:2000kx}, \texttt{FORMCalc} 
\cite{Hahn:1998yk} and \texttt{Package-X} \cite{Patel:2016fam}
and checked the result in Feynman gauge against \cite{Denner:1991kt}. 
We adopted the standard on-shell renormalization scheme for 
the electroweak parameters, consisting of $m_W$, $m_Z$ and 
the QED coupling $\alpha_{\rm{OS}}(m_Z)$, 
since the dominant scale of the Sommerfeld effect is 
the electroweak scale. We further checked 
that as $r\ll 1/m_W$, the correction coincides with 
the one-loop Coulomb potential in the massless theory 
after switching to the $\overline{\rm MS}$ renormalization 
scheme for the couplings. As a final check, we confirm 
the previously known expression for the singlet 
Yukawa potential in a Higgsed 
SU(2) theory \cite{Laine:1999rv,Schroder:1999sg} 
by taking the limit $m_W \to m_Z$ and hence $s_W \to 0, 
c_W \to 1$. 
More precisely, we confirmed the non-renormalized potential 
(Eq.~16 in \cite{Laine:1999rv}) analytically. The renormalized 
result was not 
compared, as the renormalization scheme was not fully specified.

\section{NLO potential}
\label{sec:potential}

\subsection{Result}

We obtain an analytic expression for the one-loop wino potential in 
momentum space. The Fourier transform (\ref{eq:fourier}) to the 
coordinate space potential is performed analytically where 
possible, however, for a few of the momentum-space 
functions at the one-loop order, we did not find the Fourier 
transform in a closed form, and leave it as  
a one-dimensional integral. The momentum-space 
potential is a lengthy expression, which will be given elsewhere, 
together with the potentials for the charged and spin-triplet 
channels required for relic density computations. Instead we 
provide a handy fitting function for the coordinate-space 
potential in the ${}^1S_0$ channel for charge-zero 
wino-wino scattering, which corrects (\ref{eq:VLO}) by 
\begin{equation}
\delta V(r) =    \left(\begin{array}{cc}\displaystyle 
0 &  \sqrt{2}\, \delta V_{(00)\to(+-)}
\\
 \sqrt{2}\, \delta V_{(00)\to(+-)} &
\delta V_{(+-) \to (+-)}  
\end{array}\right),
 \label{eq:VNLO}
\end{equation}
and can be easily implemented in numerical Sommerfeld codes. 
We note that the potential in the neutral channel 
$\chi^0\chi^0\to\chi^0\chi^0$ vanishes, because the only two contributing 
one-loop diagrams, the box and the crossed box diagram, cancel 
each other.

\begin{figure}[t]
\centering
\includegraphics[scale=0.45]{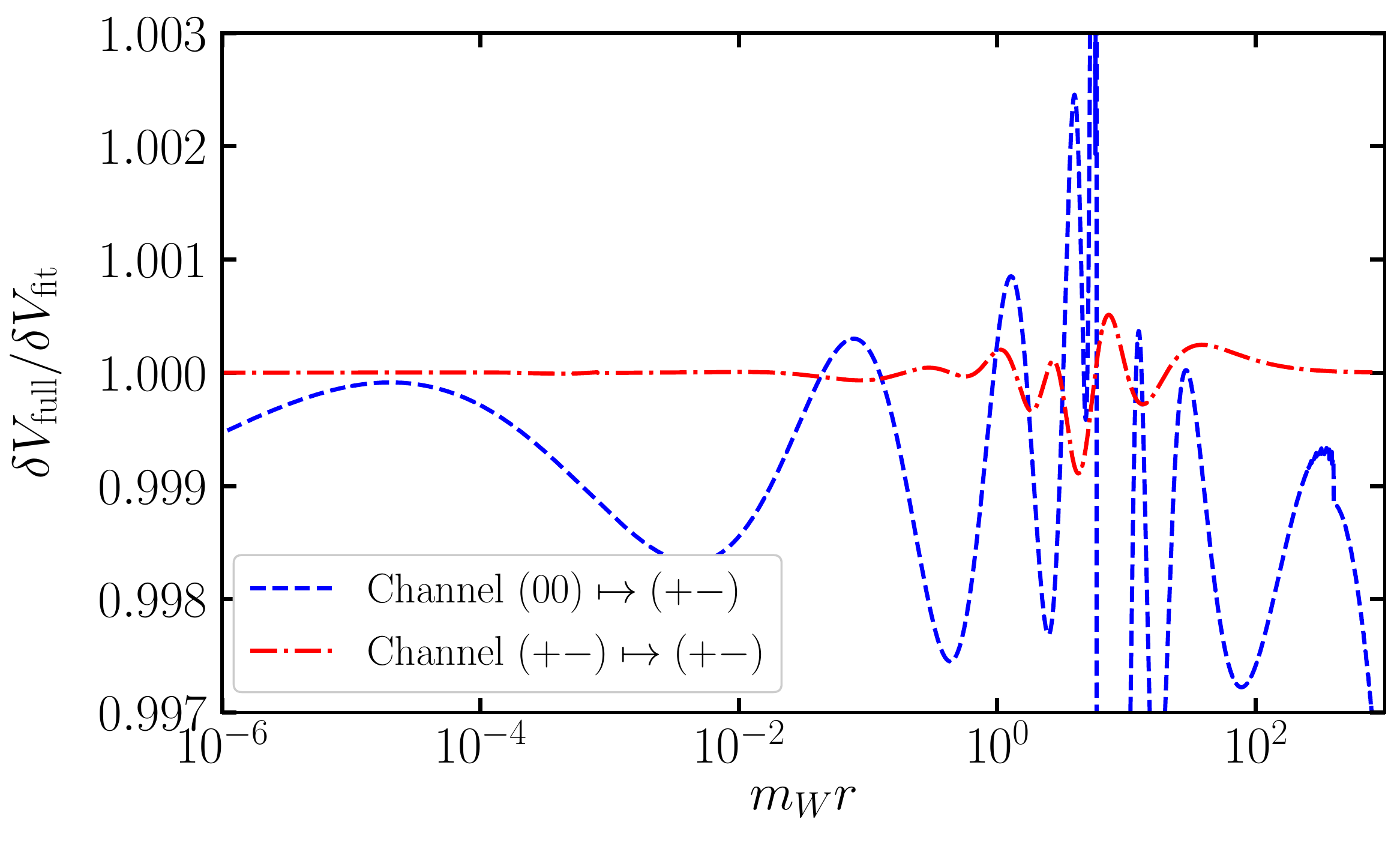}
\caption{Ratio of the numerical Fourier transform of the 
potential correction to the fitting function in the variable 
$x= m_W r$ for the channel $(00) \to (+-)$ (blue/dashed) and 
$(+-) \to (+-)$ (red/dot-dashed). The relative difference  
is in the permille range.}
\label{fig:ratiofit}
\end{figure}

\paragraph{Fitting function in the charged channel}
We use $x=m_W r$, and define
\begin{align}
\delta V^{\text{fit}}_{(+-) \to (+-)} &= 
\frac{\delta V^{r \to \infty}_{(+-) \to (+-)} }{1+ \frac{32}{11 }x^{-\frac{22}{9}}} + \frac{\delta V^{r \to 0}_{(+-)\to(+-)}}{1+\frac{7}{59} x^{\frac{61}{29}}} + \frac{\alpha}{r} \left[\frac{-\frac{1}{30} + \frac{4}{135} \log x}{1 + \frac{58}{79} x^{-\frac{17}{15}}+\frac{1}{30} x^{\frac{119}{120}} + \frac{8}{177} x^{\frac{17}{8}}}\right].
\end{align}
The fitting function is constructed from the asymptotic behaviours 
\begin{eqnarray}
\delta V^{r \to 0}_{\chi_+ \chi_- \to \chi_+ \chi_-}(r) &=& 
\frac{\alpha_2^2}{2 \pi r} \left(-\beta_{0,\text{SU(2)}}\, 
\ln(m_W r) + \frac{1960}{433}\right),
\label{eq:smallrlimit}\\
\delta V^{r \to \infty}_{\chi_+ \chi_- \to \chi_+ \chi_-}(r) &=& 
\frac{\alpha^2}{2\pi r} \,(-\beta_{0,\rm{em}})\left(\gamma_E + 
\ln(m_Z r)\right)
\label{eq:largerlimit}
\end{eqnarray}
at large and small distances and an interpolating term. 
The coefficients are rationalized to provide a compact 
expression, including the constant term $\frac{1960}{433}$ in 
(\ref{eq:smallrlimit}). $\beta_{0,\text{SU(2)}} = 19/6$ and 
$\beta_{0,\text{em}} = -80/9$ denote the leading-order coefficients  
of the beta-functions of the SU(2) and electromagnetic couplings, 
and $\gamma_E=0.577215\ldots$ is Euler's constant. 
The fitting function approximates the result of the partially 
numerical Fourier transform to better than $0.1\%$ over 
the entire distance region of interest, as shown in 
Figure~\ref{fig:ratiofit}. 

The rationalized coefficients of the numerical fitting function 
are given for the following parameters: the on-shell 
electromagnetic coupling $\alpha\equiv 
\alpha_{\rm{OS}}(m_Z)=1/128.943$ 
at the $Z$-boson mass scale, and the gauge-boson masses 
$m_W =80.385\, \rm{GeV}$ and $m_Z=91.1876\,\rm{GeV}$.  The 
cosine of the Weinberg angle and the SU(2) coupling are then 
determined from $c_W = m_W/m_Z$ and $\alpha_2 = 
\alpha_{\rm{OS}}(m_Z)/s_W^2=0.0347935$. We also 
need the top quark and Higgs boson mass, for which we take the 
on-shell masses $m_t=173.1\,\rm{GeV}$ and $m_h=125\,\rm{GeV}$. 
These parameters will also be used in the following 
discussion. For the calculation of the Sommerfeld enhancement 
below, we need in addition the two-loop mass splitting 
$\delta m_\chi = 164.1 \, \rm{MeV}$ between the 
charged and the neutral component of the wino multiplet. The 
dependence of the results on the uncertainties in these 
parameters is small enough to be ignored, except for the 
top-quark mass, as will be briefly discussed below.

\paragraph{Fitting function in the off-diagonal $(00) \to (+-)$ 
channel}
Because the correction to the potential changes sign in this 
channel near  $x_0=m_W r_0 =\frac{555 }{94}$, we did not manage 
with a single fitting function. Instead we use the piecewise 
expression 
\begin{align}
\delta V^{\text{fit}}_{(00)\to(+-)} &=
\frac{2595 \alpha^2_2}{\pi r} \times
\left\{\begin{array}{r}
\text{exp} \left[ -\frac{79 \left(L-\frac{787}{12}\right) \left(L-\frac{736}{373}\right)\left(L-\frac{116}{65}\right) \left(L^2-\frac{286L}{59}+\frac{533}{77}\right)}{34 \left(L-\frac{512}{19}\right)\left(L-\frac{339}{176}\right) \left(L-\frac{501}{281}\right)\left(L^2-\frac{268 L}{61}+\frac{38}{7}\right)}\right] ,\quad  x<x_0
\\[0.5cm]
   -\text{exp}\left[-\frac{13267 \left(L-\frac{76}{43}\right) \left(L-\frac{28}{17}\right)\left(L+\frac{37}{30}\right) \left(L^2-\frac{389L}{88}+\frac{676}{129}\right)}{5 \left(L-\frac{191}{108}\right)\left(L-\frac{256}{153}\right) \left(L+\frac{8412}{13}\right) \left(L^2-\frac{457 L}{103}+\frac{773}{146}\right)}\right] ,\quad  x>x_0 
        \end{array}\right.
\end{align}
with $L=\ln x= \ln(m_W r)$. Figure~\ref{fig:ratiofit} shows that 
the quality of the fitting function is at the few permille level, 
slightly worse than in the charged channel. At small $r$, one 
can also use the asymptotic behaviour  
$\delta V^{r \to 0}_{\chi_0 \chi_0 \to \chi_+ \chi_-}(r) = 
\delta V^{r \to 0}_{\chi_+ \chi_- \to \chi_+ \chi_-}(r)$. 

\subsection{Discussion}

\begin{figure}[t]
  \centering
  \hspace*{-0.6cm}
  \includegraphics[width=0.675\textwidth]{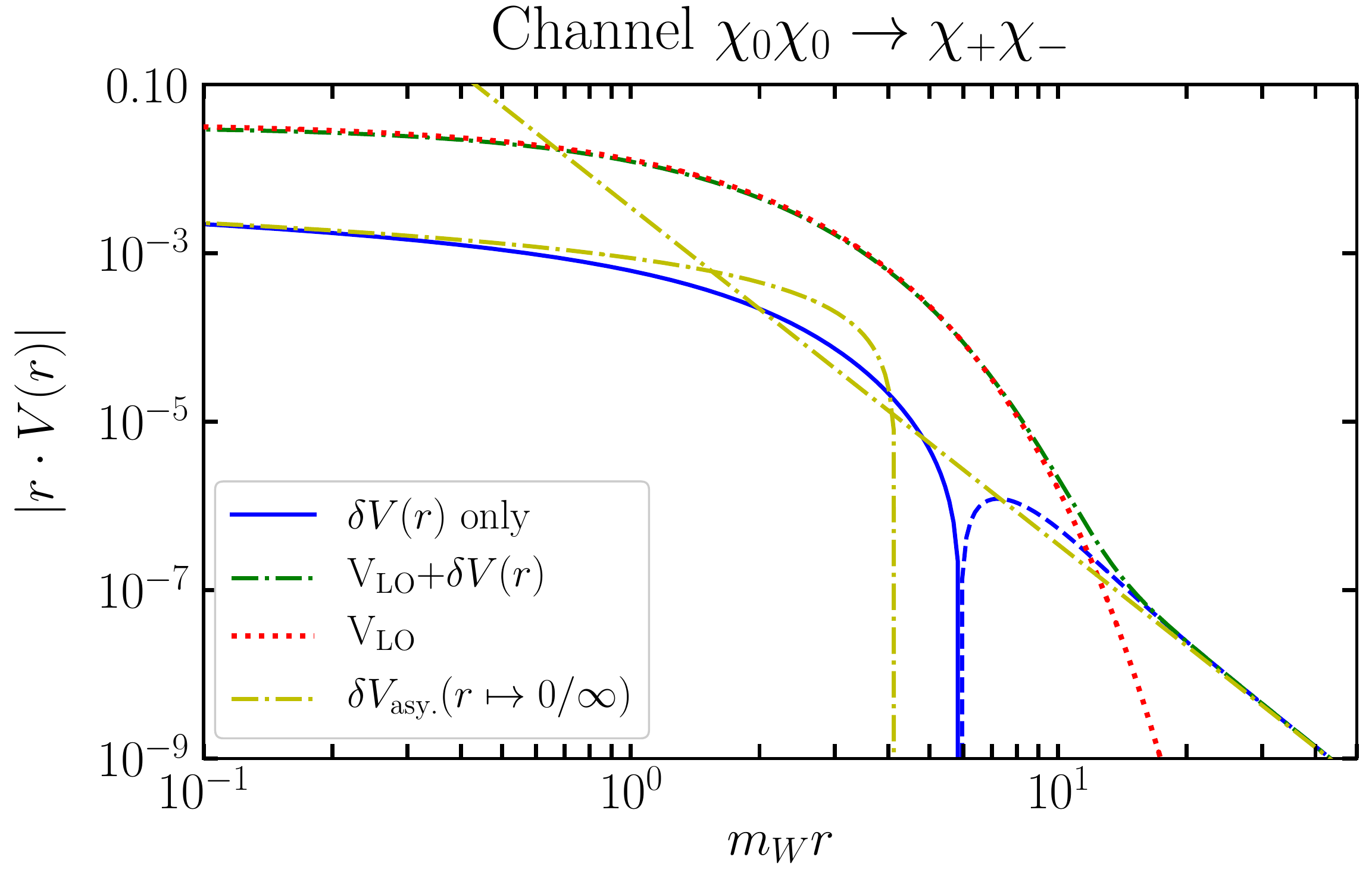}
  \includegraphics[width=0.70\textwidth]{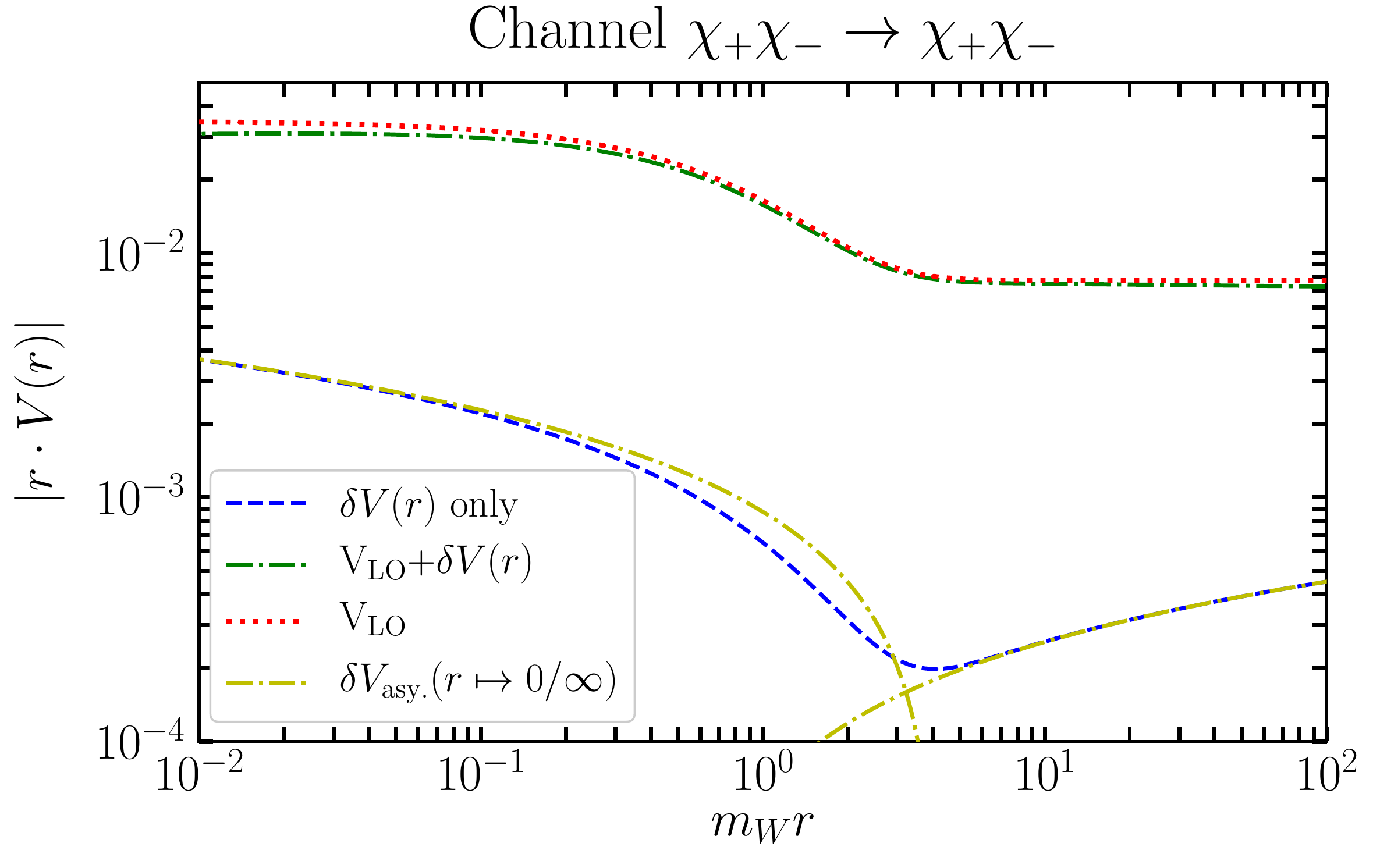}
\caption{The absolute value of LO and the NLO potential, and of 
the one-loop correction $\delta V(r)$ 
together with its asymptotic behaviours, 
all multiplied by $r$, in the 
off-diagonal and charged wino-wino scattering channel.
\label{fig:mainpotentialresult}}
\end{figure}

The following discussion of the one-loop corrected wino potential 
is based on the exact calculation and does not use the fitting 
functions from above.

The LO and NLO potential, and the NLO correction $\delta V(r)$ 
are shown in Figure~\ref{fig:mainpotentialresult} for the 
off-diagonal and charged wino-wino scattering channel. At small 
distances, the one-loop correction is governed by the correction 
(\ref{eq:smallrlimit}) to the Coulomb potential of the unbroken 
SU(2) force, which amounts to about minus $\mathcal{O}(5$-$10\%)$
for $10^{-2} < m_W r <1$ relative to the LO potential. 
At even smaller $r$, the logarithmic 
growth of the correction, see  (\ref{eq:smallrlimit}), can be 
absorbed by using a running SU(2) coupling, rather than the 
on-shell coupling. 
The one-loop term $\delta V(r)$ in the off-diagonal $\chi^0\chi^0 \to 
\chi^+\chi^-$ scattering channel (upper panel in the 
Figure) turns from positive to  negative for 
$m_W r \geq 6$ and its absolute value exceeds the tree-level 
potential at large $r$. Contrary to the naive expectation, the 
large-$r$ asymptotics of the correction is not of the Yukawa 
form $e^{-m_W r}/r$. This can be understood from the fact that 
the self-energy diagram in Figure~(\ref{fig:1looppmpm}) 
probes the transverse gauge-boson self energy $\Pi_W(-\bm{k}^2)$ at 
$\bm{k}^2\ll m_W^2$ in the large-$r$ limit. Expanding the 
self-energy resummed gauge-boson propagator 
$1/(\bm{k}^2+m_{W,0}^2-\Pi_W(-\bm{k}^2)+\delta m_W^2)$, 
where $m_{W,0}$ denotes the bare $W$ mass and $\delta m_W^2$ 
the on-shell counterterm, around $\bm{k}^2=0$, and 
transforming to coordinate space, we obtain the power-like 
rather than exponential asymptotic 
behaviour 
\begin{eqnarray}
\delta V^{r \to \infty}_{\chi_0 \chi_0 \to \chi_+ \chi_-}(r) &=& 
-\frac{9 \alpha_2^2}{\pi m_W^4 r^5}\,,
\label{eq:offdiagonallargerlimit}
\end{eqnarray}
which describes the tail of the NLO potential in the 
$\chi^0\chi^0\to\chi^+\chi^-$ scattering channel well 
for $m_W r>20$.\footnote{We assume that all fermions of the 
SM, except for the top quark, are massless.}

The behaviour of the charged scattering channel (lower panel 
in Figure~\ref{fig:mainpotentialresult}) at large distances 
is simpler, since the asymptotic behaviour becomes again 
Coulombic due to the dominance of massless photon exchange 
over the exponentially decaying terms generated by diagrams 
with $W$ and $Z$ exchange. Except in an intermediate region 
around $m_W r\sim 1$, the potential is described well by 
the asymptotic expressions (\ref{eq:smallrlimit}), 
(\ref{eq:largerlimit}). The correction is around $-4\%$ at 
$m_W r=10$, and grows logarithmically with the QED beta-function 
generated by the massless fermions of the SM.

\section{Sommerfeld effect and annihilation cross 
section}
\label{sec:xsection}

We calculate the Sommerfeld effect at NLO by solving the 
Schr\"odinger equation with the NLO wino potential employing 
the variable phase method described in \cite{Beneke:2014gja}. 
To display the NLO effect from the potential, we calculate 
the semi-inclusive $\chi^0\chi^0$ annihilation cross section 
into $\gamma+X$ with the same tree-level 
approximation\footnote{See 
\cite{Baumgart:2017nsr,Beneke:2018ssm,Beneke:2019vhz} for 
radiative corrections and Sudakov resummation of this 
annihilation rate.} 
\begin{equation}
\Gamma = 2 \,\Gamma_{\gamma\gamma}+\Gamma_{\gamma Z} 
=\frac{2\pi \alpha_2^2}{\mchi^2} 
\left(\begin{array}{cc}\displaystyle 
0 & 0 \\
0 & s_W^2 
\end{array}\right)
\end{equation}
to the short-distance annihilation matrix. 

\begin{figure}[t]
  \centering
  \hspace*{-0.6cm}
  \includegraphics[width=0.8\textwidth]{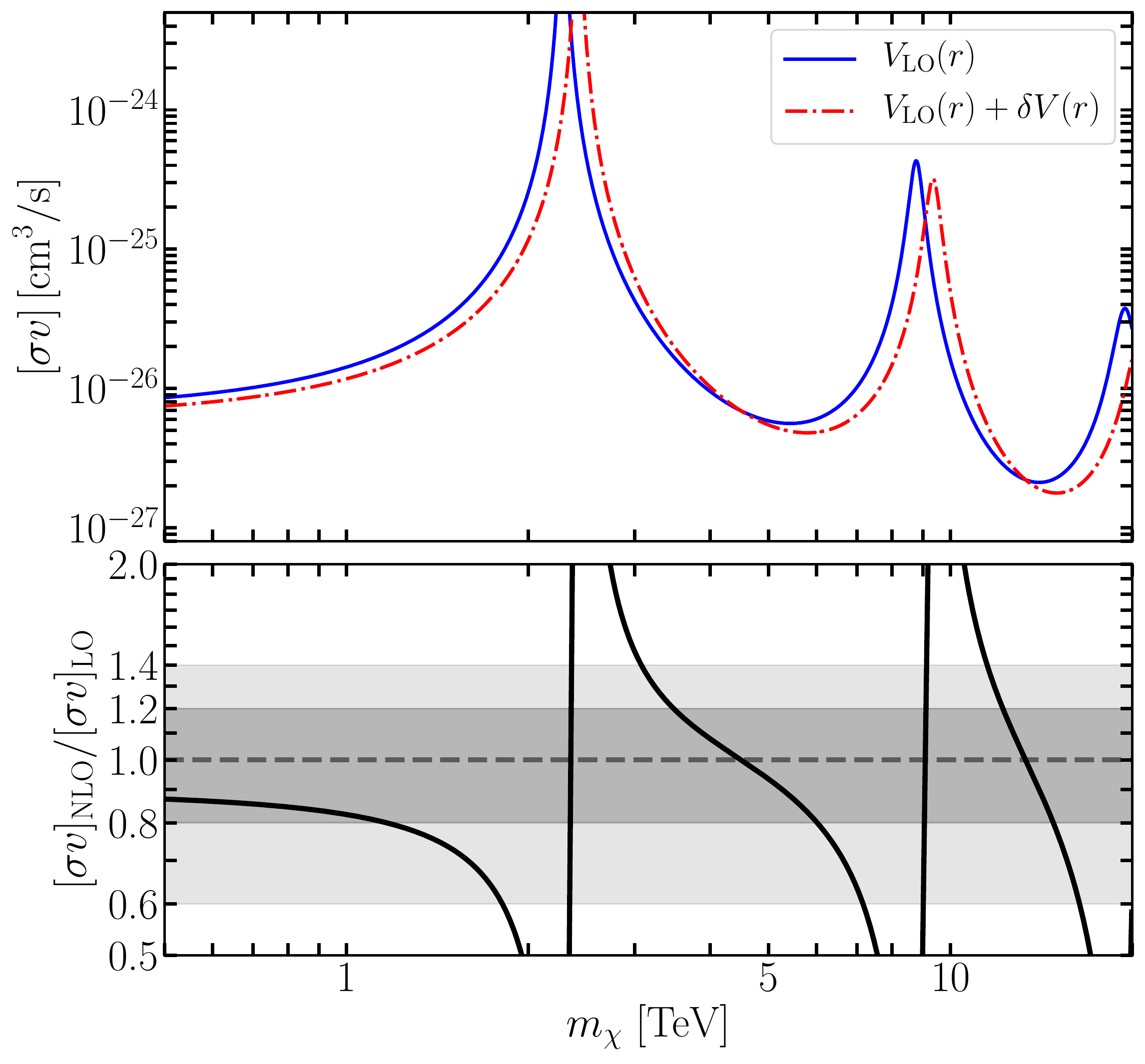}\caption{$\sigma v$ calculated with the LO (solid/blue) 
and the NLO (dash-dotted/red) potential. The lower panel shows the 
ratio of the NLO to LO result with dark (light) grey bands  
to visualize the range where the correction stays below 
20\% (40\%).
\label{fig:mainSFresult}}
\end{figure}

In the upper panel of Figure~\ref{fig:mainSFresult}
we show $\sigma v$, the annihilation 
cross section times velocity calculated with the LO (solid/blue) 
and the NLO (dash-dotted/red) potential in the mass range 
$\mchi = 0.5\ldots 20$~TeV for the DM particle, which 
covers the onset of the Sommerfeld enhancement at small 
masses and the first two resonances. We recall that the 
observed relic density is achieved for a wino mass of 
$2.88$~TeV \cite{Beneke:2016ync}. That the NLO correction is visible on a 
logarithmic plot already indicates that it is significant. 
The location of the first two Sommerfeld resonances shifts 
from $2.283$ ($8.773$) TeV at LO to $2.419$ ($9.355$) TeV at NLO. 
Since the resonances both move to larger masses, the 
NLO correction changes sign in the mass range between 
the resonances and always remains sizeable. This can 
be seen in the subtended lower panel of Figure~\ref{fig:mainSFresult}, 
which displays the ratio of the NLO to LO annihilation 
cross section. The ratio evidently blows up near the 
resonances due to the location-shift, but it is larger 
than 20\% for wide mass ranges, and always larger than 
the typical 3\% for an electroweak loop correction.

\begin{figure}[t]
\centering
\includegraphics[scale=0.45]{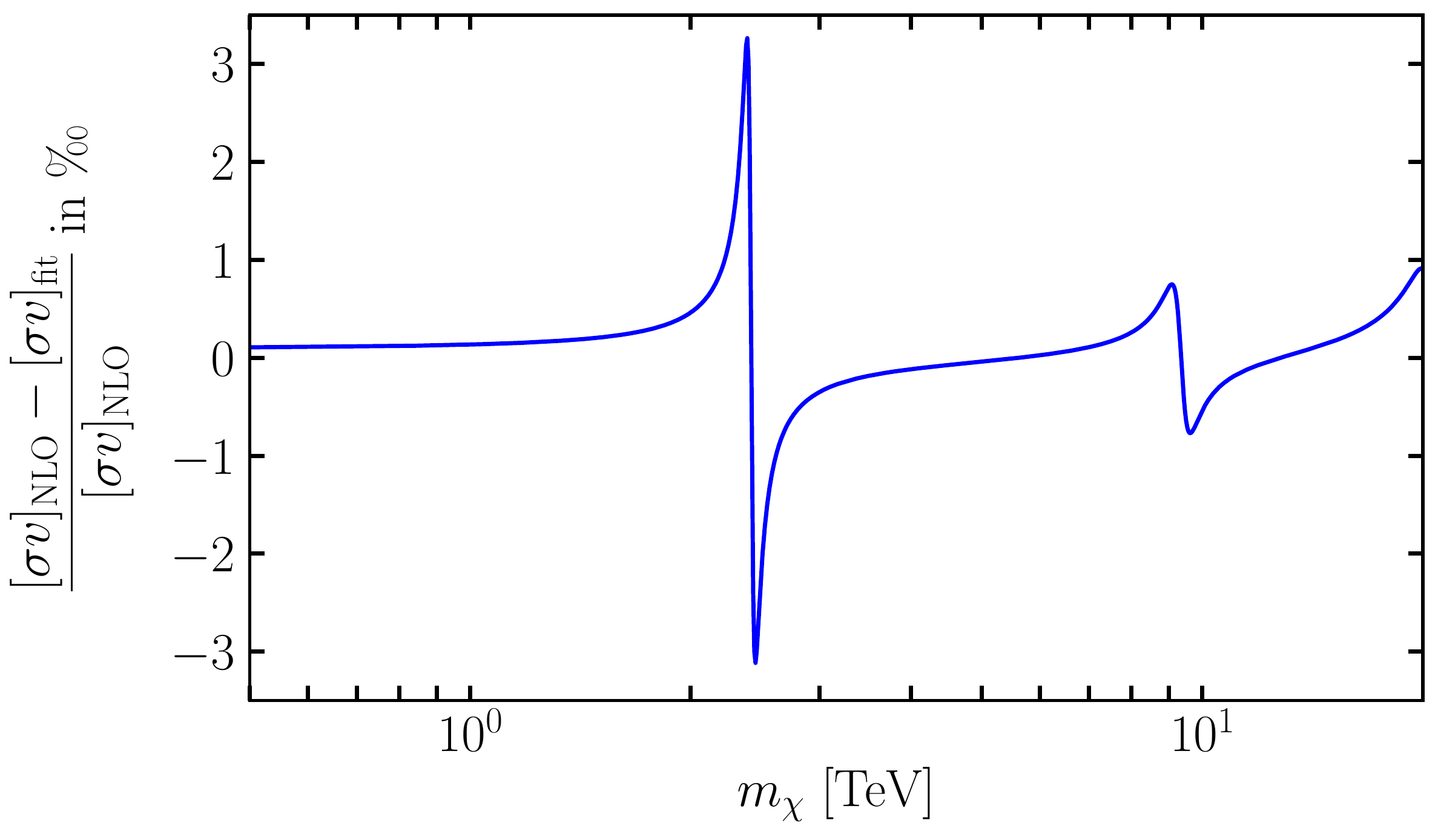}
\caption{Relative error in permille of the Sommerfeld enhanced cross section using the full NLO potential vs. the fitting function for 
the relevant range of  values of DM mass~$m_\chi$.}
\label{fig:SEratiofit}
\end{figure}

For completeness, we show in Figure~\ref{fig:SEratiofit} the 
accuracy of the annihilation cross section when instead of the 
exact computation of the NLO potential, the fitting functions are 
used. The error is at most 0.3\% near the first resonance and 
usually substantially smaller. The first (second) resonance 
position changes by only 0.1 GeV (0.2 GeV). 

The above results depend on the value of the top quark mass 
through the gauge boson self energies. We adopted the on-shell 
mass, since the characteristic scale for the Sommerfeld 
effect is the electroweak scale. If instead we choose the 
$\overline{\rm MS}$ mass $\overline{m}_t(\overline{m}_t) = 
163.35$~GeV, the NLO resonances are located at 
$2.408$ TeV, $9.311$~TeV, respectively. This amounts to a 
change of about 8\% in the size of the shifts from LO to NLO. 
The overall picture remains unaffected.

In summary, we computed the NLO correction to the wino potential. 
We find that the Sommerfeld  resonances are shifted by about 
6\% to larger values, from 2.283~TeV to 2.419~TeV for the first 
resonance, and find sizeable corrections over the
entire mass range relevant for wino-like DM. This effect is 
generally larger than a typical electroweak loop correction and 
should be included in precision predictions of annihilation rates 
in the wino model, such 
as \cite{Ovanesyan:2016vkk,Beneke:2018ssm,Beneke:2019vhz}.
Furthermore, the size of the effect suggests further investigation 
of its relevance for the relic DM abundance, which requires the 
calculation of the NLO potentials in all coannihilation 
channels.

\paragraph{Acknowledgements} This work was supported in part 
by the DFG Collaborative Research Centre ``Neutrinos and 
Dark Matter in Astro- and Particle Physics''  (SFB 1258).

\bibliography{dm}

\end{document}